\documentclass[prd,aps]{revtex4}
\usepackage{graphicx,latexsym}

\begin{document}

\title{Consistent and mimetic discretizations in general
relativity} 

\author{Cayetano Di Bartolo$^{1}$, Rodolfo Gambini$^{2}$,
and Jorge Pullin$^{3}$} 

\affiliation { 1. Departamento de F\'{\i}sica,
Universidad Sim\'on Bol\'{\i}var,\\ Aptdo. 89000, Caracas 1080-A,
Venezuela.\\ 2. Instituto de F\'{\i}sica, Facultad de Ciencias, Igu\'a
4225, esq. Mataojo, Montevideo, Uruguay. \\ 3. Department of Physics
and Astronomy, Louisiana State University, Baton Rouge, LA 70803-4001}

\date{May 30th 2004}

\begin{abstract}
A discretization of a continuum theory with constraints or conserved
quantities is called {\it mimetic} if it mirrors the conserved laws or
constraints of the continuum theory at the discrete level. Such
discretizations have been found useful in continuum mechanics and in
electromagnetism.  We have recently introduced a new technique for
discretizing constrained theories.  The technique yields
discretizations that are consistent, in the sense that the constraints
and evolution equations can be solved simultaneously, but it cannot be
considered mimetic since it achieves consistency by determining the
Lagrange multipliers. In this paper we would like to
show that when applied to general relativity linearized around a
Minkowski background the technique yields a discretization that is
mimetic in the traditional sense of the word.  We show this
using the traditional metric variables and also the Ashtekar new
variables, but in the latter case we restrict ourselves to the
Euclidean case.  We also argue that there appear to exist conceptual
difficulties to the construction of a mimetic formulation of the full
Einstein equations, and suggest that the new discretization scheme can
provide an alternative that is nevertheless close in spirit to the
traditional mimetic formulations. 
\end{abstract}

\maketitle

\section{Introduction}

Continuum theories, either mechanical systems or field theories,
usually have conservation laws and sometimes constraints. When one
discretizes the equations of these theories, for instance in order to
solve them numerically on a computer, or for ``quantization on the
lattice'' purposes, the resulting discrete equations will usually fail
to preserve the conserved quantities of the continuum theory upon
evolution. Similar comments apply to constraints. Although one may
have discrete equations resulting from discretizing the constraints of
the continuum theory, if one chooses initial data that solves these
equations exactly, they will fail to be solved upon discrete
evolution.

{\em Mimetic} discretizations are discretizations of continuum
theories that preserve conserved quantities or constraints in the
discrete theory that {\em mimic} those of the continuum theory.  There
is quite a body of literature \cite{mimetic} on mimetic
discretizations in the context of continuum mechanics and
electromagnetism. The literature on Hamiltonian lattice QCD implicitly
considers a mimetic discretization of Yang--Mills theory, although
this fact is not usually emphasized.

Some authors have considered the question of whether mimetic
discretizations of general relativity can be constructed
\cite{private,Me}. It is well known that if one discretizes the
Einstein equations, the Hamiltonian and momentum constraint, which
should hold for all time if satisfied initially (ignoring for the
moment the issue of spatial boundaries), fail to do so in the discrete
theory. Although there has been success in generating mimetic
formulations of linearized relativity, it appears unlikely that
something similar will be available for the full theory (or even for
the linearized theory on non-trivial backgrounds or slicings).  This
is due to the fact that discretized derivatives fail to satisfy
Leibnitz' rule and therefore the nonlinear terms when discretized do
not have properties that mirror those of the continuum \cite{Me}.

We have recently introduced a new approach to the discretization of
theories, particularly of theories with constraints
\cite{DiGaPu,GaPuprl} called ``consistent discretization''. The
technique guarantees that the resulting discrete equations are
compatible, i.e., they admit a common set of solutions (something that
is not generically true if one discretizes the equations of a
constrained theory).  The technique has been tried out in the context
of cosmological solutions of the Einstein equations \cite{cosmo}, of
BF theory and of Maxwell and Yang--Mills theories on the lattice
\cite{DiGaPu}.  Current investigations are testing it for the Gowdy
models.

In this paper we would like to show that the technique we proposed,
when applied to the Einstein theory linearized around a Minkowski
background yields a discrete formulation that is mimetic. That is, the
discretized constraints are exactly preserved under evolution without
determining the Lagrange multipliers. We first consider linearized
general relativity in terms of the traditional metric variables. We
then consider it in terms of Ashtekar's variables, which have the
advantage of being closer to the discretizations used in Yang--Mills
theories (although in this case we restrict ourselves to Euclidean
general relativity).

In the consistent discretization scheme, equations are discretized
with variables evaluated at two (or more) different levels in time.
This includes the constraints of general relativity, which in the
discrete theory therefore can only be viewed as ``pseudo'' constraints
(we will reserve the word constraint for expressions that involve all
variables evaluated at the same instant of time, as in traditional
canonical terminology). These equations, with variables discretized at
mixed instants of time are the equations that are solved by the
consistent discretization scheme. Of course, if one is in a regime in
which the time-step is small, then satisfying the pseudo-constraints
implies that the usual discrete constraints (with all the variables at
the same time-step) are approximately satisfied as well.  Therefore the
resulting scheme cannot be strictly called mimetic, although it
approximately is. We will show that if one uses a discretization for
general relativity that is mimetic in the linearized case, one further
improves the accuracy with which the consistent scheme for the full
non-linear theory satisfies the constraints. This encourages further
studies of these discretization schemes in the context of numerical
applications.

In the next section we will present a brief summary of the 
consistent discretizations scheme. In the following two sections
we apply it to linearized gravity, first with the traditional
variables and then with the Ashtekar variables. We end with a 
discussion and proposals for further research.

\section{Consistent discretization of constrained theories}

We illustrate the technique with a mechanical system for simplicity,
but there is no problem working it out for field theories, since upon
discretization the latter become mechanical systems. We assume we
start from an action in the continuum, written in first-order form,
\begin{equation}
S=\int L(q,p) dt
\end{equation}
with 
\begin{equation}
L(q,p)=p\, \dot{q} -H(q,p) -\lambda \phi(q,p)
\end{equation}
where $\lambda$ is a Lagrange multiplier and the theory has a single
(it is immediate to incorporate several) constraint $\phi(q,p)=0$. The
discretization of the action yields $S=\sum_{0}^N L(n,n+1)$, where
\begin{equation}
L(n,n+1)= p_n (q_{n+1}-q_n) -\epsilon H(q_n,p_n)-\lambda_{n}
\phi(q_n,p_n),
\end{equation}
where $\epsilon=t_{n+1}-t_n$ and we have absorbed an $\epsilon$ in the
definition of the Lagrange multipliers.

We will now view the Lagrangian as the generator of a type 1 canonical
transformation between the instant $n$ and the instant $n+1$. In
ordinary classical mechanics parlance, given a canonical
transformation between a canonical pair $q,p$ and a new canonical pair
$Q,P$, the generating function of a type 1 canonical transformation is
a function of $q,Q$, $F(q,Q)$ and the canonically conjugate momenta
are defined by $P=\partial F /\partial Q$, $p=\partial F/ \partial q$.
In our case we will view $q_n,p_n,\lambda_n$ and
$q_{n+1},p_{n+1},\lambda_{n+1}$ as ``configuration variables'' and
will assign to each of them a canonically conjugate momentum through
the canonical transformation,
\begin{eqnarray}
P^q_{n+1} &=& {\partial L(n,n+1) \over \partial q_{n+1}} ,\\
P^p_{n+1}&=& {\partial L(n,n+1) \over \partial p_{n+1}},\\
P^{\lambda}_{n+1}&=& {\partial L(n,n+1) \over \partial \lambda_{(n+1)}},\\
P^q_{n} &=& -{\partial L(n,n+1) \over \partial q_{n}},\\
P^p_{n}&=& -{\partial L(n,n+1) \over \partial p_{n}},\\
P^{\lambda}_{n}&=& -{\partial L(n,n+1) \over \partial \lambda_{(n)}}.
\end{eqnarray}

If one explicitly computes the partial derivatives with the Lagrangian
given, one can eliminate the $p,P^p$ and $P^\lambda$ to yield a more 
familiar-looking set of equations,
\begin{eqnarray}
P^q_{n+1}-P^q_{n} &=& -\epsilon {\partial H(q_n,P^q_{n+1}) 
\over \partial q_n} - \lambda_{nB}
{\partial \phi^B(q_n,P^q_{n+1}) \over \partial q_n},\nonumber\\
q_{n+1}-q_n &=&
\epsilon {\partial H(q_n,P^q_{n+1}) \over \partial P^q_{n+1}} 
+ \lambda_{nB} {\partial \phi^B(q_n,P^q_{n+1}) 
\over \partial P^q_{n+1}},\nonumber\\
\phi^B(q_n,P^q_{n+1}) &=&0.\label{10}
\end{eqnarray}

These indeed look like a discrete version of equations for a system
with constraints. However, there are important differences. First of
all, notice that as an evolution system the equations are
implicit. Secondly, if one solves the first two equations one obtains
$P^q$ and $q$ as functions of the initial data and the Lagrange
multipliers. The last equation however, will generically not hold. One
will have to choose specific values for the Lagrange multipliers at
each time-step (and if one is dealing with a field theory at each
point in space) for all the equations to be solved.

Notice that there can be particular cases in which the system does not
determine the Lagrange multipliers. For instance, consider a totally
constrained system like general relativity. There the Hamiltonian
vanishes. Suppose now that the constraint in (\ref{10}) is only a
function of $q_n$. Then the evolution equation for $q_{n+1}$ implies
that $q_{n+1}=q_n$ and the constraint is automatically
preserved. Therefore the resulting formulation is mimetic in the
traditional sense of the word, a constraint that is just the discrete
version of the continuum constraint is preserved under evolution by
the discrete evolution equations. A similar situation develops if the
constraint is only a function of $P^q$.

If the Hamiltonian is non-vanishing, and the constraint depends only
on $P^q_{n+1}$ then the latter is not automatically preserved upon
evolution, but it cannot be satisfied by choosing the Lagrange
multipliers either since they drop out from the relevant evolution
equations. On the other hand if the constraint is only a function of
$q_n$, its preservation could be enforced by choosing the Lagrange
multipliers (the asymmetry between $P^q$ and $q$ in this treatment
comes from the fact that we chose to write the equations as
``propagating forward'' in time, if one had chosen to propagate
backwards, the roles of $q$ and $P^q$ in this discussion would be
reversed).

Summarizing, the consistent discretization technique consists of
discretizing the action and working out the resulting equations of
motion for the discrete theory from it through the canonical
transformation that implements time evolution. The resulting evolution
equations (and constraints) are made a consistent set of nonlinear
algebraic equations by considering the Lagrange multipliers as
dynamical variables one has to solve for.  In particular situations,
the Lagrange multipliers are not determined by the equations. In such
cases the resulting set of equations and constraints has to be
consistent since it has been derived from a variational principle and
the resulting discrete theory is mimetic in the traditional sense of
the word: the constraints are automatically preserved upon evolution.
In the other case, when the Lagrange multipliers are determined
the resulting discrete theory is based on a consistent set of 
algebraic equations, but as one can see in equation (\ref{10}),
one is enforcing the constraints with some variables evaluated at
instant $n$ and some at instant $n+1$. For small stepsizes, this
implies that the constraints with all the variables evaluated at
the same instant of time are approximately preserved. The
resulting theory therefore cannot be called mimetic in the traditional
sense of the word, although it can do a good job of preserving
(approximately) the discrete constraints.

In the next two sections we apply this technique to linearized general
relativity. We will see that the resulting theories do not determine
the Lagrange multipliers, preserve the constraints automatically, and
therefore are mimetic in the traditional sense of the word. We will
not discuss the case of full general relativity here, but in several
examples we have considered elsewhere for the non-linear theory
(cosmologies \cite{cosmo}, Gowdy spacetimes) the Lagrange multipliers
are determined. Therefore it is unlikely that this method will yield a
mimetic formulation for full GR. However, as we argued above, it will
yield a formulation that approximates general relativity well in
certain regimes and in such regimes the discrete constraints are
enforced approximately very well. We believe it is likely that this is
``as close as one will get'' to a mimetic formulation of full general
relativity.

\section{Linearized general relativity in terms of metric variables}

In this section we will apply the technique we described in the
previous section to linearized general relativity written in terms
of the traditional variables. We assume the background is the Minkowski
metric.

\subsection{Continuum formulation}

We start with the Arnowitt, Deser and Misner (ADM) \cite{ADM} form of
the action of general relativity,
\begin{equation}
S= \int d^4x \left[\pi^{ab}\dot{q}_{ab}-N C -N_a C^a\right],
\end{equation}
where 
\begin{eqnarray}
C&=&{1\over \sqrt{q}}\left[\pi^{ab}\pi_{ab}-\frac{1}{2}
(\pi_b^b)^2\right]-\sqrt{q}\, {}^{(3)}R\\
C^a&=&-2 \pi^{ab}_{;b}
\end{eqnarray}
and the variables $(q_{ab},N,N_a)$ are related to the four dimensional
metric ${}^{(4)}g_{\mu\nu}$ through,
\begin{eqnarray}
q_{ab}&=&{}^{(4)}g_{ab},\\
N&=&\left(-\,{}^{(4)}g_{00}\right)^{-1/2}\\
N_a&=&{}^{(4)}g_{0a},
\end{eqnarray}

The indices $a,b,c$ run from 1 to 3. $\sqrt{q}$ is the determinant of
the spatial metric $q_{ab}$ and ${}^{(3)}R$ is its Ricci curvature
scalar.  The momenta $\pi^{ab}$ are related to the extrinsic curvature
of the space-like surfaces $x^0=t={\rm constant}$ through
$\pi^{ab}=-\sqrt{q} \left[K^{ab}-q^{ab} K^c_c\right]$ and indices are
raised and lowered with the spatial metric. The semicolon denotes
covariant differentiation with respect to the Christoffel connection
of the spatial metric. Variation with respect to
$\pi^{ab},q_{ab},N,N_a$ yields the Einstein equations.  In particular
variation of $N,N_a$ gives rise to four constraints $C=0,C^a=0$
usually referred to as (super)Hamiltonian and momentum (or
diffeomorphism) constraints.

We have chosen the ADM action since it is one of the most
traditionally used in general relativity. Modern numerical
implementations favor the use of formulations in which the evolution
equations are manifestly symmetric-hyperbolic. This is not the case
for the ADM equations. In principle there is no obstruction in
applying our technique to any action, but it just is the case that
there has been little investigations about formulating the
symmetric-hyperbolic formulations as deriving from an action
principle. This will require further study and therefore we decided to
concentrate on this paper on the ADM action for simplicity.

We now consider that the spacetime metric is given by a static
background metric plus small perturbations 
${}^{(4)} g_{\mu\nu}={}^{(4)} g^{(0)}_{\mu\nu}+h_{\mu\nu}$. For
simplicity we make the further choice that the foliation is such that
the zeroth order shift $N_a=0$ and the zeroth order extrinsic
curvature is therefore zero $\pi^{ab}=0$. The constraint equations to
leading order in the perturbations are given by \cite{moncrief},

\begin{eqnarray}
C^a &=& -2 p^{ab}{}_{;b}\label{diflin}\\
C&=&-\sqrt{q}\left[h_{ab}{}^{;ab} -h_{;a}{}^{;a} -h_{ab} {}^{(3)}
R^{ab}\right]
\label{hamlin}
\end{eqnarray}

In these expressions $p^{ab}$ is the linear portion of the canonical
momentum $\pi^{ab}$ and ${}^{(3)} R^{ab}$ is the Ricci tensor of the
background metric. The action for the linearized theory is,
\begin{equation}
S=\int d^4x \left[p^{ab} \dot{h}_{ab}-N^{(0)} H -N^{(1)}_a C^a
-N^{(1)} C \right]
\end{equation}
where we have kept track of the order in the perturbation expansion of 
the lapse and the shift (recall that we assume zero shift in the
background). The constraints $C,C^a$ are given by the expressions
above, where only terms up to order linear have been kept. The
quantity
\begin{equation}
H={1 \over \sqrt{q}} \left[p^{ab} p_{ab} -\frac{1}{2} p^2\right]
+\frac{1}{2}\sqrt{q} \left[\frac{1}{2} h_{ab;c}
h^{ab;c}-h_{ab;c}h^{ac;b}- \frac{1}{2} h_{;a}h^{;a}+2
h_{;a}h^{ab}{}_{;b}+h h^{ab}{}_{;ab}- h h_{ab} {}^{(3)}R^{ab}\right]
\end{equation}
is a true Hamiltonian density (not a constraint) that is responsible
for the evolution of the canonical variables, and is multiplied in the
action times the lapse of the background space-time.

At this point we can make an important observation. The momentum
constraint (\ref{diflin}) is only a function of the momenta $p^{ab}$
(the covariant derivative is with respect to the background metric)
and the Hamiltonian constraint (\ref{hamlin}) is only a function
of the configuration variables $q^{ab}$. Therefore, as we discussed in
section II, our discretization technique will not determine the value
of the Lagrange multipliers. The resulting theory therefore can only
either be: mimetic or inconsistent. We will proceed to show that the
resulting discrete theory is indeed consistent.

\subsection{Discretization}

We start by discretizing the linearized action, $S=\sum_{n=1}^N
L(n,n+1)$, where,
\begin{eqnarray}
L(n,n+1)&=&\sum_{\vec{m}}\left(\sum_{a,b=1}^3 \left\{p_{ab}(n,\vec{m})
\left(h_{ab}(n+1,m)-h_{ab}(n,m)\right)\right.\right. \\ &&-
N(n,\vec{m}) \left[ h_{ab}(n,\vec{m}+\vec{e}_a+\vec{e}_b)
-h_{ab}(n,\vec{m}-\vec{e}_a+\vec{e}_b)\right.\nonumber\\
&&\left.-h_{ab}(n,\vec{m}+\vec{e}_a-\vec{e}_b)
+h_{ab}(n,\vec{m}-\vec{e}_a-\vec{e}_b)\right.\nonumber\\
&&-\left.h_{aa}(n,\vec{m}+2\vec{e}_b) +
2h_{aa}(n,\vec{m})-h_{aa}(n,\vec{m}-2\vec{e}_b)\right] \nonumber\\
&&\left.\left.-N_a(n,\vec{m})\left[2 p_{ab}(n,\vec{m}+\vec{e}_b)- 2
p_{ab}(n,\vec{m}-\vec{e}_b)\right]\right\}-H(n,\vec{m})
\begin{array}{c}
\\
\\
\\
\end{array}
\right)\nonumber
\end{eqnarray}
and,
\begin{eqnarray}
H(n,\vec{m})&=& \sum_{a,b=1}^3\left[ p_{ab}(n,\vec{m})^2
-\frac{1}{2}p_{aa}(n,\vec{m})p_{bb}(n,\vec{m})\right]\\
&&+\frac{1}{2}\sum_{a,b,c=1}^3
\left[\frac{1}{2}\left(h_{ab}(n,\vec{m}+\vec{e}_c)
-h_{ab}(n,\vec{m}-\vec{e}_c)\right)^2\right.\nonumber\\
&&- \left(h_{ab}(n,\vec{m}+\vec{e}_c)
-h_{ab}(n,\vec{m}-\vec{e}_c)\right) \left(h_{ac}(n,\vec{m}+\vec{e}_b)
-h_{ac}(n,\vec{m}-\vec{e}_b)\right)\nonumber\\ && -\frac{1}{2}
\left(h_{aa}(n,\vec{m}+\vec{e}_c)
-h_{aa}(n,\vec{m}-\vec{e}_c)\right)\left(h_{bb}(n,\vec{m}
+\vec{e}_c)-h_{bb}(n,\vec{m}-\vec{e}_c)\right)\nonumber\\
&&+ 2
\left(h_{cc}(n,\vec{m}+\vec{e}_a)-h_{cc}(n,\vec{m}-\vec{e}_a)\right)
\left(h_{ab}(n,\vec{m}+\vec{e}_b)-h_{ab}(n,\vec{m}-\vec{e}_b)\right)\nonumber\\
&&+h_{cc}(n,\vec{m})\left[ h_{ab}(n,\vec{m}+\vec{e}_a+\vec{e}_b)
-h_{ab}(n,\vec{m}+\vec{e}_a-\vec{e}_b)\right.\nonumber\\
&&\left.\left.
+h_{ab}(n,\vec{m}-\vec{e}_a-\vec{e}_b)
-h_{ab}(n,\vec{m}-\vec{e}_a+\vec{e}_b)\right]\frac{}{}\right],\nonumber
\end{eqnarray}
where we have assumed that the background metric is Minkowski and
we have chosen the zeroth order lapse equal to unity and we have
dropped the $(1)$ superscript from the first order lapse and 
shift. We have also chosen a centered prescription for spatial 
derivatives, with the following conventions,
i.e. $\phi(i)_{,x}=\phi(i+1)-\phi(i-1)$ and 
$\phi(i)_{,xx}=\phi(i+2)+\phi(i-2)-2\phi(i)$ and similarly for
higher derivatives. This choice of 
prescription is needed for two reasons: i) it ensures that
``summation by parts'' (ignoring boundaries) is satisfied, which
is important when taking variations of the action; ii) it makes
the successive application of two first derivatives the second
derivative, etc. This is important when proving mimetism.

The Lagrangian is the generator of the canonical transformation
that materializes evolution from instant $n$ to instant $n+1$. 
Specifically, we will introduce the canonically conjugate momenta
as we discussed in the previous section,
\begin{eqnarray}
&&P^h_{ab}(n+1,\vec{m}) = p_{ab}(n,\vec{m})\label{eq24}\\
&&P^p_{ab}(n+1,\vec{m}) = 0\label{pricons1}\\
&&P^N(n+1,m)= 0\label{pricons2}\\
&&P^N_a(n+1,m)= 0\label{pricons3}\\
&&P^h_{ab}(n,\vec{m}) = p_{ab}(n,\vec{m})
+N(n,\vec{m}-\vec{e}_a-\vec{e}_b)-
N(n,\vec{m}-\vec{e}_a+\vec{e}_b)\label{pricons4}\\
&&-N(n,\vec{m}+\vec{e}_a-\vec{e}_b)+N(n,\vec{m}+\vec{e}_a+\vec{e}_b)\nonumber\\
&&-\delta_{ab} \sum_{c=1}^3\left(N(n,\vec{m}-2\vec{e}_c)-2
N(n,\vec{m})+N(n,\vec{m}+2\vec{e}_c)\right)\nonumber\\
&&+\frac{1}{2}\sum_{c=1}^3 \left[\left(h_{ab}(n,\vec{m})-
h_{ab}(n,\vec{m}-2\vec{e}_c)\right)-\left(
h_{ab}(n,\vec{m}+2\vec{e}_c)-h_{ab}(n,\vec{m})\right)\right]\nonumber\\
&&-\frac{1}{2}\sum_{c=1}^3 \left[
\left(h_{ac}(n,\vec{m}+\vec{e}_b-\vec{e}_c)
-h_{ac}(n,\vec{m}-\vec{e}_b-\vec{e}_c)\right)
-\left(h_{ac}(n,\vec{m}+\vec{e}_b+\vec{e}_c)-h_{ac}(n,\vec{m}-\vec{e}_b
+\vec{e}_c)\right)\right]
\nonumber\\
&&-\frac{1}{2}\sum_{c=1}^3 \left[
\left(h_{bc}(n,\vec{m}+\vec{e}_a-\vec{e}_c)
-h_{bc}(n,\vec{m}-\vec{e}_a-\vec{e}_c)\right)
-\left(h_{bc}(n,\vec{m}+\vec{e}_a+\vec{e}_c)-h_{bc}(n,\vec{m}-\vec{e}_a
+\vec{e}_c)\right)\right]
\nonumber\\
&&-\frac{1}{2}\delta_{ab}\sum_{c,d=1}^3 \left[\left(
h_{dd}(n,\vec{m})-h_{dd}(n,\vec{m}-2\vec{e}_c)\right)
-
\left(h_{dd}(n,\vec{m}+2\vec{e}_c)
-h_{dd}(n,\vec{m})\right)\right]\nonumber\\
&&+{\delta_{ab}\over 2} \sum_{c,d=1}^3 \left[
h_{cd}(n,\vec{m}+\vec{e}_d-\vec{e}_c)-
h_{cd}(n,\vec{m}-\vec{e}_d-\vec{e}_c)
-h_{cd}(n,\vec{m}+\vec{e}_c+\vec{e}_d)+
h_{cd}(n,\vec{m}+\vec{e}_c-\vec{e}_d)\right]\nonumber\\
&&+\frac{1}{2} \sum_{c=1}^3 \left[\left(
h_{cc}(n,\vec{m}+\vec{e}_a-\vec{e}_b)-
h_{cc}(n,\vec{m}-\vec{e}_a-\vec{e}_b)\right)
-\left(h_{cc}(n,\vec{m}+\vec{e}_a+\vec{e}_b)-
h_{cc}(n,\vec{m}-\vec{e}_a+\vec{e}_b)\right)\right],\nonumber\\
&&P^p_{ab}(n,\vec{m}) =
-\left(h_{ab}(n+1,\vec{m})-h_{ab}(n,\vec{m})\right)+2
p_{ab}(n,\vec{m})\label{pricons1n}\\
&&-\sum_{c=1}^3 p_{cc}(n,\vec{m})\delta_{ab}+
 N_a(n,\vec{m}-\vec{e}_b)- N_a(n,\vec{m}+\vec{e}_b)
+ N_b(n,\vec{m}-\vec{e}_a)- N_b(n,\vec{m}+\vec{e}_a)\nonumber\\
&&P^{N_a}(n,m)= \sum_{b=1}^3\left[ 2 p_{ab}(n,\vec{m}+\vec{e}_b)-
2 p_{ab}(n,\vec{m}-\vec{e}_b)\right]\label{pricons2n}\\
&&P^N(n,m)= \sum_{a,b=1}^3 \left[
h_{ab}(n,\vec{m}+\vec{e}_a+\vec{e}_b)
-h_{ab}(n,\vec{m}-\vec{e}_a+\vec{e}_b)\right.\label{pricons3n}
\\
&&\left. -h_{ab}(n,\vec{m}+\vec{e}_a-\vec{e}_b)
+h_{ab}(n,\vec{m}-\vec{e}_a-\vec{e}_b)
-h_{aa}(n,\vec{m}+2\vec{e}_b)+
2h_{aa}(n,\vec{m})-h_{aa}(n,\vec{m}-2\vec{e}_b)\right]\nonumber
\end{eqnarray}

The system has four primary constraints 
(\ref{pricons1}-\ref{pricons4}). Preserving these constraints in time
implies, via (\ref{pricons1n}-\ref{pricons3n}) that the linearized
Hamiltonian and momentum constraints are satisfied,
\begin{eqnarray}
C_a&=&2 \sum_{b=1}^3 \left[P^h_{ab}(n,\vec{m}+\vec{e}_b)-
 P^h_{ab}(n,\vec{m}-\vec{e}_b)\right]=0\label{discmom}\\
C&=&
\sum_{a,b=1}^3 \left[
h_{ab}(n,\vec{m}+\vec{e}_a+\vec{e}_b)
-h_{ab}(n,\vec{m}-\vec{e}_a+\vec{e}_b) -h_{ab}(n,\vec{m}+\vec{e}_a-\vec{e}_b)
+h_{ab}(n,\vec{m}-\vec{e}_a-\vec{e}_b)\right.\nonumber\\
&&\left.-h_{aa}(n,\vec{m}+2\vec{e}_b)+
2h_{aa}(n,\vec{m})-h_{aa}(n,\vec{m}-2\vec{e}_b)\right]=0
\label{discham}
\end{eqnarray}
Constraints (\ref{pricons1},\ref{pricons4}) can be imposed strongly,
the second constraint determines the variable $p_{ab}$.  This
eliminates the variable $p_{ab}$ and its canonically conjugate momenta
from the theory.

We now combine (\ref{eq24}) and (\ref{pricons4}) to get the evolution
equation for $P^h$,
\begin{eqnarray}
&&P^h_{ab}(n+1,\vec{m}) =
P^h_{ab}(n,\vec{m}) \label{ecuevolp}\\
&&-N(n,\vec{m}-\vec{e}_a-\vec{e}_b)+
N(n,\vec{m}-\vec{e}_a+\vec{e}_b)+N(n,\vec{m}+\vec{e}_a-\vec{e}_b)
-N(n,\vec{m}+\vec{e}_a+\vec{e}_b)
\nonumber \\
&&+\delta_{ab} \sum_{c=1}^3\left(N(n,\vec{m}-2\vec{e}_c)-2
N(n,\vec{m})+N(n,\vec{m}+2\vec{e}_c)\right)\nonumber\\
&&-\frac{1}{2}\sum_{c=1}^3 \left[\left(h_{ab}(n,\vec{m})-
h_{ab}(n,\vec{m}-2\vec{e}_c)\right)-\left(
h_{ab}(n,\vec{m}+2\vec{e}_c)-h_{ab}(n,\vec{m})\right)\right]\nonumber\\
&&+\frac{1}{2}\sum_{c=1}^3 \left[
\left(h_{ac}(n,\vec{m}+\vec{e}_b-\vec{e}_c)
-h_{ac}(n,\vec{m}-\vec{e}_b-\vec{e}_c)\right)
-\left(h_{ac}(n,\vec{m}+\vec{e}_b+\vec{e}_c)-h_{ac}(n,\vec{m}-\vec{e}_b+
\vec{e}_c)\right)\right]
\nonumber\\
&&+\frac{1}{2}\sum_{c=1}^3 \left[
\left(h_{bc}(n,\vec{m}+\vec{e}_a-\vec{e}_c)
-h_{bc}(n,\vec{m}-\vec{e}_a-\vec{e}_c)\right)
-\left(h_{bc}(n,\vec{m}+\vec{e}_a+\vec{e}_c)-h_{bc}(n,\vec{m}-\vec{e}_a+
\vec{e}_c)\right)\right]
\nonumber\\
&&+\frac{1}{2}\delta_{ab}\sum_{c,d=1}^3 \left[\left(
h_{dd}(n,\vec{m})-h_{dd}(n,\vec{m}-2\vec{e}_c)\right)
-
\left(h_{dd}(n,\vec{m}+2\vec{e}_c)
-h_{dd}(n,\vec{m})\right)\right]\nonumber\\
&&-{ \delta_{ab}\over 2} \sum_{c,d=1}^3 \left[\left(
h_{cd}(n,\vec{m}+\vec{e}_d-\vec{e}_c)-
h_{cd}(n,\vec{m}-\vec{e}_d-\vec{e}_c)\right)
-\left(h_{cd}(n,\vec{m}+\vec{e}_c+\vec{e}_d)-
h_{cd}(n,\vec{m}+\vec{e}_c-\vec{e}_d)\right)\right]\nonumber\\
&&-\frac{1}{2} \sum_{c=1}^3 \left[\left(
h_{cc}(n,\vec{m}+\vec{e}_a-\vec{e}_b)-
h_{cc}(n,\vec{m}-\vec{e}_a-\vec{e}_b)\right)
-\left(h_{cc}(n,\vec{m}+\vec{e}_a+\vec{e}_b)-
h_{cc}(n,\vec{m}-\vec{e}_a+\vec{e}_b)\right)\right],\nonumber
\end{eqnarray}
and from (\ref{pricons1n}) we get the evolution equation for $h$, 
\begin{eqnarray}
&&h_{ab}(n+1,\vec{m})=
h_{ab}(n,\vec{m})+2
P^h_{ab}(n,\vec{m}) -\delta_{ab}\sum_{f=1}^3
P^h_{ff}(n,\vec{m}) \label{ecuevolh}\\
&&+
N_a(n,\vec{m}-\vec{e}_b)- N_a(n,\vec{m}+\vec{e}_b)
+N_b(n,\vec{m}-\vec{e}_a)- N_b(n,\vec{m}+\vec{e}_a)\nonumber\\
&&-2N(n,\vec{m}-\vec{e}_a-\vec{e}_b)+
2N(n,\vec{m}-\vec{e}_a+\vec{e}_b)
+2N(n,\vec{m}+\vec{e}_a-\vec{e}_b)-2N(n,\vec{m}+\vec{e}_a+\vec{e}_b)
\nonumber\\
&&-\sum_{c=1}^3 \left[2 h_{ab}(n,\vec{m})-
h_{ab}(n,\vec{m}-2\vec{e}_c)-
h_{ab}(n,\vec{m}+2\vec{e}_c)\right]\nonumber\\
&&+\sum_{c=1}^3 \left[
h_{ac}(n,\vec{m}+\vec{e}_b-\vec{e}_c)
-h_{ac}(n,\vec{m}-\vec{e}_b-\vec{e}_c)
-h_{ac}(n,\vec{m}+\vec{e}_b+\vec{e}_c)
+h_{ac}(n,\vec{m}-\vec{e}_b+\vec{e}_c)\right]
\nonumber\\
&&+\sum_{c=1}^3 \left[
h_{bc}(n,\vec{m}+\vec{e}_a-\vec{e}_c)
-h_{bc}(n,\vec{m}-\vec{e}_a-\vec{e}_c)
-h_{bc}(n,\vec{m}+\vec{e}_a+\vec{e}_c)
+h_{bc}(n,\vec{m}-\vec{e}_a+\vec{e}_c)\right]
\nonumber\\
&&+ \frac{1}{2}\delta_{ab}\sum_{c,d=1}^3 \left[
2 h_{dd}(n,\vec{m})-h_{dd}(n,\vec{m}-2\vec{e}_c)
-
h_{dd}(n,\vec{m}+2\vec{e}_c)
\right]\nonumber\\
&&- \frac{1}{2}\delta_{ab} \sum_{c,d=1}^3 \left[
h_{cd}(n,\vec{m}+\vec{e}_d-\vec{e}_c)-
h_{cd}(n,\vec{m}-\vec{e}_d-\vec{e}_c)
-h_{cd}(n,\vec{m}+\vec{e}_c+\vec{e}_d)+
h_{cd}(n,\vec{m}+\vec{e}_c-\vec{e}_d)\right]\nonumber\\
&&-\sum_{c=1}^3 \left[
h_{cc}(n,\vec{m}+\vec{e}_a-\vec{e}_b)-
h_{cc}(n,\vec{m}-\vec{e}_a-\vec{e}_b)
-h_{cc}(n,\vec{m}+\vec{e}_a+\vec{e}_b)+
h_{cc}(n,\vec{m}-\vec{e}_a+\vec{e}_b)\right],\nonumber\\
\nonumber
\end{eqnarray}

A first point to be noted is that the evolution equations have
resulted in an explicit evolution scheme. This is usually not the
case, it is a particularity of the linearized theory that the
evolution is explicit. It should be noted that the evolution 
equations obtained are just a straightforward discretization of
the evolution equations one would obtain in the continuum 
by working out the variations of the continuum action.

We have checked, using a computer algebra code, that the evolution
equations (\ref{ecuevolh},\ref{ecuevolp}) exactly preserve the
constraints (\ref{discmom},\ref{discham}), or more precisely that,
\begin{eqnarray}
C_a(n+1,m)&=&C_a(n,m),\\
C(n+1,m) &=& C(n,m) +\sum_{a=1}^3
\left[C_a(n,m+\vec{e}_a)-C_a(n,m-\vec{e}_a)\right].
\end{eqnarray}

This result was expected since we used differentiation operators that
ensure that mixed discrete spatial derivatives commute, and that one
can integrate by parts (more precisely ``sum by parts''), and that is
all that is needed in a linear theory on a Minkowski background to
show that the constraints are preserved upon evolution. It is
interesting to compare this result with that of Meier \cite{Me}. He
finds a mimetic discretization of linearized general relativity around
Minkowski spacetime, but using staggered grids. This is a natural
approach, for instance, in electromagnetism and Yang--Mills theory
(and it is the one we will take in the next section where we deal with
gravity with the Ashtekar variables).

It would be interesting to generalize these results to the case of
linearization around a static background. In that case it is not
obvious that the formulation would result automatically mimetic.  In
fact, the failure of the Leibnitz rule at a discrete level implies
that it will be difficult to find a mimetic formulation since the
equations now will have non-constant coefficients and one will need
Leibnitz' rule to show conservation.  Our formalism will yield a
consistent formulation, but it is possible that it will require
determining the Lagrange multipliers.  

\subsection{Stability}

We have discretized the time derivatives without centering them (that
is, we have used a stencil that is first order accurate only). The
reason for this is that the canonical theory is much cleaner with only
two levels in time involved in the derivatives. It is possible to use
derivatives that are second order accurate in time and use our
construction by rewriting the theory in terms new variables in such a
way that the resulting theory has derivatives that are first order
accurate, but we will not do this here.

The spatial derivatives, on the other hand, were centered (this was
required in order to have summation by parts). The resulting scheme
is therefore ``forward in time centered in space'', a recipe that is
not stable, for instance, for the advection or the wave equation. We
therefore would like to check if our scheme is stable. To simplify
things, we will consider (\ref{ecuevolh},\ref{ecuevolp}) and make
the following assumptions: the metric and extrinsic curvatures are
diagonal and only depend on the coordinates $t,x$, the lapse is 
unity and the shift is zero. The resulting equations therefore are,
\begin{equation}
P^h_{11}(n+1,\vec{m}) =
P^h_{11}(n,\vec{m}) 
-\frac{1}{2} \sum_{c=2}^3 \left[
2h_{cc}(n,\vec{m})-
h_{cc}(n,\vec{m}-2\vec{e}_1)
-h_{cc}(n,\vec{m}+2\vec{e}_1)\right]
,\nonumber
\end{equation}

\begin{equation}
P^h_{22}(n+1,\vec{m}) =
P^h_{22}(n,\vec{m}) 
+\frac{1}{2} 
\left[2h_{33}(n,\vec{m})-h_{33}(n,\vec{m}-2\vec{e}_1)
-h_{33}(n,\vec{m}+2\vec{e}_1)
\right]
\end{equation}

\begin{equation}
P^h_{33}(n+1,\vec{m}) =
P^h_{33}(n,\vec{m}) 
+\frac{1}{2} 
\left[2h_{22}(n,\vec{m})-h_{22}(n,\vec{m}-2\vec{e}_1)
-h_{22}(n,\vec{m}+2\vec{e}_1)
\right]
\end{equation}

\begin{eqnarray}
h_{11}(n+1,\vec{m})&=&
h_{11}(n,\vec{m})+2
P^h_{11}(n,\vec{m}) -\sum_{f=1}^3
P^h_{ff}(n,\vec{m}) \\
&&- \frac{1}{2}\sum_{d=2}^3 \left[
2 h_{dd}(n,\vec{m})-h_{dd}(n,\vec{m}-2\vec{e}_1)
-
h_{dd}(n,\vec{m}+2\vec{e}_1)
\right]\nonumber
\end{eqnarray}

\begin{eqnarray}
h_{22}(n+1,\vec{m})&=&
h_{22}(n,\vec{m})+2
P^h_{22}(n,\vec{m}) -\sum_{f=1}^3
P^h_{ff}(n,\vec{m}) \\
&&-\frac{1}{2}\left[2 h_{22}(n,\vec{m})-
h_{22}(n,\vec{m}-2\vec{e}_1)-
h_{22}(n,\vec{m}+2\vec{e}_1)\right]\nonumber\\
&&+ \frac{1}{2} \left[
2 h_{33}(n,\vec{m})-h_{33}(n,\vec{m}-2\vec{e}_1)
-
h_{33}(n,\vec{m}+2\vec{e}_2)
\right]\nonumber
\end{eqnarray}

\begin{eqnarray}
h_{33}(n+1,\vec{m})&=&
h_{33}(n,\vec{m})+2
P^h_{33}(n,\vec{m}) -\sum_{f=1}^3
P^h_{ff}(n,\vec{m}) \\
&&-\frac{1}{2}\left[2 h_{33}(n,\vec{m})-
h_{33}(n,\vec{m}-2\vec{e}_1)-
h_{33}(n,\vec{m}+2\vec{e}_1)\right]\nonumber\\
&&+ \frac{1}{2} \left[
2 h_{22}(n,\vec{m})-h_{22}(n,\vec{m}-2\vec{e}_1)
-
h_{22}(n,\vec{m}+2\vec{e}_2)
\right]\nonumber
\end{eqnarray}

As a test case, we concentrate on a subfamily of solutions of the
equations, in which $h_{11}=P_{11}=0$ and $h_{22}=-h_{33}$
and $P_{22}=-P_{33}$. In that case, the equations reduce to,
\begin{equation}
P^h_{22}(n+1,\vec{m}) =
P^h_{22}(n,\vec{m}) 
-\frac{1}{2} 
\left[2h_{22}(n,\vec{m})-h_{22}(n,\vec{m}-2\vec{e}_1)
-h_{22}(n,\vec{m}+2\vec{e}_1)
\right],
\end{equation}
\begin{equation}
h_{22}(n+1,\vec{m})=
h_{22}(n,\vec{m})+2
P^h_{22}(n,\vec{m}) 
-\left[2 h_{22}(n,\vec{m})-
h_{22}(n,\vec{m}-2\vec{e}_1)-
h_{22}(n,\vec{m}+2\vec{e}_1)\right].
\end{equation}

We have performed a Von Neumann analysis of
this system and confirmed that the scheme is stable provided the
Courant factor is less than one. So at least for this particular
subcase the scheme is stable. A more complete analysis is needed to
guarantee stability in general.

\section{Linearized general relativity in terms of Ashtekar variables}
\subsection{Continuum formulation}
We will now apply the technique we outlined in the previous section to
general relativity linearized around Minkowski space using the
Ashtekar formulation. The formulation of linearized gravity with the
new variables was first discussed by Ashtekar and Lee
\cite{Ashtekar:wp}. The discussion presented in that paper required
the use of complex variables if one was to describe general relativity
with metrics with a Lorentzian signature (alternatively, one could
consider real variables, but then the theory described the Euclidean
signature sector.)  Developments that have taken place in the field
since the publication of that paper that allow to consider the
Lorentzian sector using real variables \cite{Ba,Th}, but we will see
that the discretized theory is more problematic in this case and we
will not discuss it in detail in this paper.

The Ashtekar canonical variables consist of a set of triads with
density weight 1, $E^{ai}$ and a (complex) $SO(3)$ connection
$A_{ai}$. In this notation $a,b,\ldots$ are spatial vector indices and
$i,j,\ldots$ range from 1 to 3. Following Ashtekar and Lee we omit
using tildes to denote density weights since in this context they do
not play an important role. To linearize the theory around Minkowski,
we choose a fixed background $(E^{ai}=E^{ai}_0,A_{ai}=0)$ in the phase
space and consider fluctuations around it. In Cartesian coordinates,
$E^{ai}_0=\delta^{ai}$. The triad is therefore given by,
\begin{equation}
E^{ai}=\delta^{ai}+e^{ai},
\end{equation}
and therefore the background metric has components
$q^{ab}=\delta^{ab}$ and its determinant is unity and therefore
density weights are all trivial. We will denote by $A_{ai}$ the
fluctuations of the connection. The Poisson bracket of the canonical
variables is $\{e^{ai}(x),A_{bj}(y)\}=i\delta^a_b \delta^i_j
\delta(x-y)$.

The Ashtekar formulation has, in addition to the usual diffeomorphism
and Hamiltonian constraints of the metric canonical formulation of
general relativity, a set of additional constraints that make the
formulation invariant under triad rotations. The additional
constraints take the form of a Gauss law, which linearized will read,
\begin{equation}
{\cal G}^i_L = \partial_a e^{ai} + \epsilon^{ija} A_{aj} =0,
\end{equation}
where from now on the subscript $L$ means we have kept the minimum
required number of terms in the perturbative expansion. In spite of
the second term, one can check that if one computes the Poisson
bracket of two Gauss laws, they commute, that is, they form an Abelian
algebra. The internal symmetry group of the linearized theory is
therefore $U(1)^3$.

Ignoring boundary terms, the (super)Hamiltonian for general relativity
can be written as,
\begin{equation}
H = \int d^3x \left[N E^a_i E^b_j \left(F_{ab}^k \epsilon_{ijk} 
-{(\beta^2-\sigma)\over \beta^2} (\Gamma_a^i-\sigma\beta A_a^i)(\Gamma_b^j-
\sigma\beta A_b^j)\right)
+N^a E^a_i F_{ab}^i,\right]\label{super}
\end{equation}
and in the full theory it vanishes identically. The parameter $\beta$
is called the Immirzi parameter and the parameter $\sigma$ is equal to
$+1$ for the Euclidean case and $-1$ for the Lorentzian
signature. Classically, different values of the Immirzi parameter correspond
to different representations of the same theory. The quantities
$\Gamma_a^i$ are the spin connections compatible with the triads,
defined by,
\begin{equation}
\partial_{[a} \bar{E}_{b]}^i+\epsilon^i_{jk} 
\Gamma^j_{[a} \bar{E}_{b]}^k =0,
\end{equation}
where the $\bar{E}$'s are the triads (without density weight), related
to the Ashtekar variables by ${E}^a_i={\rm det}(\bar{E}) \bar{E}^a_i$,
or equivalently, $E^a_i =\bar{E}^j_b\bar{E}^k_c \epsilon^{abc}
\epsilon_{ijk}$. Indices are lowered and raised with the flat
Euclidean metric. One can obtain an explicit expression for the spin 
connection in terms of the triads,
\begin{equation}
\Gamma_c^i = \epsilon^{ab}_c (\partial_a e_b^i-\partial_a {\rm Tr}(e) 
\delta_b^i).
\end{equation}

To study it in the linearized theory, we need to
choose a lapse and a shift.  The natural choice is to use as zeroth
order lapse and shifts the ones that would preserve the spatial
background metric explicitly time-independent. This corresponds to a
lapse $N=1$ and a shift $N^a=0$. So we will write $N_L=1+\nu$ and
$N^a_L=\nu^a$, and these will become Lagrange multipliers in the
linearized theory.  The super-Hamiltonian then separates into two
pieces, one that acts as a Hamiltonian and another piece that is given
by the Lagrange multipliers times constraints of the linearized
theory. These constraints are,
\begin{eqnarray}
C^L_a&=& -i f_{ab}^b=0,\\
C^L&=& -i \epsilon^{ab}_c f^c_{ab}=0,
\end{eqnarray}
where $f_{ab}^i = 2\partial_{[a} A_{b]}^i$ is the linearized field
strength, and the first one is the linearized momentum constraint and
the latter the linearized Hamiltonian constraint. The non-vanishing
Hamiltonian for the linearized theory is given by,
\begin{eqnarray}
H_L &=& \int d^3x \left(2\epsilon^{ib}_k f_{ab}^k e^a_i 
+ (A_a^a A_b^b -A_a^b A_b^a)\right.\\
&&\left.-{\beta^2-\sigma \over \beta^2}\left[ (\Gamma_a^a-\sigma\beta A_a^a)
(\Gamma_a^b-\sigma\beta A_a^b))
-(\Gamma_a^b-\sigma\beta A_a^b)(\Gamma_b^a-\sigma\beta A_b^a)\right]\right)
\nonumber
\end{eqnarray}

\subsection{Discretizing the full theory on the lattice}

In this section we review some results of reference
\cite{DiGaPu,GaPuprl} where we discretized general relativity on
the lattice. In the next section we will particularize these
results to the case of linearized general relativity.  We start by
considering an action for general relativity written in terms of
Ashtekar's variables (see for instance \cite{CDJM} and the book by
Ashtekar \cite{Asbook} page 47),
\begin{equation}
L=\int E^{ai} F_{a0}^i - H
\end{equation}
where $N$ and $N^a$ are the lapse and shift and $H$ the
super-Hamiltonian (\ref{super}). We will particularize to the
Euclidean case $\sigma=1$ and choose the Immirzi parameter
$\beta=1$ which correspond to the original form of Ashtekar's
variables, for simplicity (see section VI for more details).
From now on we will not assume Einstein's summation convention
and present the summations explicitly, since many expressions
would otherwise be confusing.  The
Lagrangian can be discretized as follows,
\begin{eqnarray}
L(n,n+1) &=& -\frac{1}{4}\sum_v {\rm Tr}\left[\sum_a E^a_{n,v}
(h^{a0}_{n,v} -h^{0a}_{n,v}) - \sum_{a,b}K^{ab}_{n,v}
(h^{ab}_{n,v}- h^{ba}_{n,v}) \right.
 \nonumber
\\
&&+\left.\sum_a\alpha_{a,n,v} \left(h^a_{n,v}
\left(h^a_{n,v}\right)^\dagger -\mbox{\large 1}\right)+
\beta_{n,v} \left(h^0_{n,v}
\left(h^0_{n,v}\right)^\dagger-\mbox{\large
1}\right)\right]\label{disclag}
\end{eqnarray}
where $h^a_{n+1}$ represents an holonomy along the $a$ direction
at instant $n+1$, $h^0_n$ represents the ``vertical'' (time-like)
holonomy. The holonomy associated with a plaquette in the 
$\alpha\beta$ ($\alpha\neq \beta$) plane ($\alpha,\beta=0\ldots3$) is
\begin{equation}
h^{\alpha\beta}_{n,v} \equiv h^\alpha_{n,v}
h^\beta_{n,v+e_\alpha} (h^\alpha_{n,v+e_\beta})^\dagger
(h^\beta_{n,v})^\dagger \,,
\end{equation}
and
\begin{equation}
K^{ab}_{n,v} \equiv \frac{1}{2} \left[ (E^{a}_{n,v}
E^{b}_{n,v}-E^{b}_{n,v} E^{a}_{n,v}) N_{n,v} + N^a_{n,v}
E^{b}_{n,v}-N^b_{n,v} E^{a}_{n,v}\right]\,.
\end{equation}
We will
assume that the holonomies are matrices of the form $h=\sum_I h^I T^I$
where $T^0=I$ and $T^a=-i\sigma^a$ where $\sigma^a$ are the Pauli
matrices. The indices $n,v$ represent a label for ``time'' $n$ and
a spatial label for the vertices of the lattice $v$. The
elementary unit vectors along the spatial directions are labeled
as $e_a$, so for instance $n+e_1$ labels the nearest neighbor to
$n$ along the $e_1$ direction. The unit vector in the timelike
direction is 
$e_0$ and we chose $h^\alpha_{n,v+e_0}
\equiv h^\alpha_{n+1,v}$. The quantities $E^a_{n,v}$ are elements
of the algebra of $su(2)$ and $\alpha$ and $\beta$ are Lagrange
multipliers, the last two terms of the Lagrangian enforcing the
condition that the holonomies are elements of $SU(2)$. We use the
usual conventions of lattice gauge theories in which one has
oriented links and the natural variables are the holonomies in a
given orientation and based at a given vertex.  If we need to
traverse back, as in the case of closed loops one then considers
the adjoint of the holonomy based at the vertex one is ending at.

The discretization of the field tensor is based on,
\begin{equation}
F^i_{a b} \rightarrow -\frac{1}{4}\mbox{Tr}\left[ (h^{a b}_{n,v} -
h^{b a}_{n,v} )T^i\right] \,.
\end{equation}

Instead of working out the equations of motion for this action, we
will, in the next section, particularize it to the linearized case
and work out the relevant equations of motion, which is equivalent
to working the equations first and then linearizing if appropriate
perturbative orders are kept.

\subsection{The linearized theory on the lattice}

We now proceed to linearize the action. We start with the
holonomies. The explicit form of the linearized holonomy is
\begin{equation}
h^{\alpha}_{v} = \mbox{\large 1} + \sum_i \phi^{\alpha i}_{v}T^i
\end{equation}
where we have dropped the subscript $n$ we used in the last
section to indicate the time level in order to make the notation
more compact (but we will make it explicit when things are
evaluated at $n+1$). In this equation $\phi^{\alpha i}_{v}T^i$ is
an element of the algebra that can be viewed as a ``phase'' (it
corresponds to the logarithm of the path-ordered exponential of
the connection along the direction $\alpha$). The holonomy of a 
plaquette in the plane $\alpha\beta$ is (neglecting higher order terms), 
\begin{equation}\label{hPlaqueta}
h^{\alpha \beta}_{v} = h^\alpha_{v} h^\beta_{v+e_\alpha}
(h^\alpha_{v+e_\beta})^\dagger (h^\beta_{v})^\dagger
 = (1 - \sum_i\Phi_{2v}^{\alpha\beta\, ii}) \mbox{\large 1}
+\sum_i(\Phi_{1v}^{\alpha\beta i} +\Phi_{2v}^{\alpha\beta i}) T^i.
\end{equation}

The first order contribution is,
\begin{equation}\label{FasePlaqueta}
\Phi_{1v}^{\alpha\beta k} \equiv
 + \phi^{\alpha k}_v
 - \phi^{\beta k}_v
 - \phi^{\alpha k}_{v+\hat{e}_\beta}
 + \phi^{\beta k}_{v+\hat{e}_\alpha}  \,,
\end{equation}
and the second order contribution is,
\begin{eqnarray}
\Phi_{2v}^{\alpha\beta\, ij} &\equiv&
 - \phi^{\alpha i}_v \phi^{\beta j}_v
 + \phi^{\alpha i}_v \phi^{\beta j}_{v+\hat{e}_\alpha}
 + \phi^{\alpha i}_{v+\hat{e}_\beta} \phi^{\beta j}_v
 - \phi^{\alpha j}_{v+\hat{e}_\beta} \phi^{\beta i}_{v+\hat{e}_\alpha}
 - \phi^{\alpha i}_v \phi^{\alpha j}_{v+\hat{e}_\beta}
 - \phi^{\beta j}_v \phi^{\beta i}_{v+\hat{e}_\alpha}
\\
\Phi_{2v}^{\alpha\beta\, k} &\equiv& \sum_{ij} \epsilon^{ijk}
\Phi_{2v}^{\alpha\beta\, ij}\,.
\end{eqnarray}

We now linearize the expression for $K$ defined in the previous
section, by noting that to first order,
\begin{equation}
 e^a_v =   \sum_i (\delta^{ai} + e^{ai}_v) T^i
\end{equation}
and ignoring higher order terms we get that
\begin{equation}\label{vectorK}
K^{ab}_{v} = \sum_i ( \epsilon^{abi} +K_{1,v}^{abi}
 )T^i\,.
\end{equation}
with
\begin{equation}
K_1^{abk} \equiv \epsilon^{abk} \nu_v + \frac{1}{2}(\nu_v^a
\delta^{bk} - \nu_v^b \delta^{ak})+ \sum_i (e^{ai}_v
\epsilon^{ibk} -e^{bi}_v \epsilon^{iak}).
\end{equation}

We now consider the first term in the discretized Lagrangian
(\ref{disclag}). Substituting the expression for the holonomy 
around a plaquette (\ref{hPlaqueta}) 
we get the following identity, valid up to second order,
\begin{equation}
-\frac{1}{4}\mbox{Tr}\left[\sum_a E^a_{n,v} (h^{a0}_{n,v}
-h^{0a}_{n,v}) \right] = \sum_a \Phi_{1v}^{a 0 a} + \sum_a
\Phi_{2v}^{a 0 a} + \sum_{ak} e^{ak}_v \Phi_{1v}^{a 0 k}
\end{equation}
and we note that when one considers the sum over all vertices,
the first term of the right hand side yields a total derivative
with respect to time that can be ignored in the Lagrangian.

For the second term in (\ref{disclag}) we use (\ref{hPlaqueta}) and
(\ref{vectorK}), getting the following identity, valid up to 
second order,
\begin{equation}
\frac{1}{4}\mbox{Tr} [\sum_{ab} K^{ab}_{n,v} (h^{ab}_{n,v}-
h^{ba}_{n,v})] = -\sum_{abk}(\epsilon^{abk}\Phi_{1v}^{a b k} +
\epsilon^{abk}\Phi_{2v}^{a b k} + K_{1v}^{abk}\Phi_{1v}^{a b k})
\,.
\end{equation}

When one considers the sum over all vertices the first term on the
right hand side of this expression vanishes.  The resulting Lagrangian
therefore can be written as,
\begin{eqnarray}
L &=& -\sum_v  \left\{
 \sum_{ai} e^{ai}_v(
 \phi^{ai}_{n+1,v}-\phi^{ai}_{n,v})
 + \sum_{ijk}\epsilon_{ijk} ( \phi^{i j}_v \phi^{ik}_{n+1,v}
 + \Phi^{ijk}_{2v} +  \nu_v  \Phi_{1v}^{ijk} )
 \right.
\nonumber
\\
&&
 + \sum_{aijk} 2 e^{ai}_v \epsilon^{ijk} \Phi_{1v}^{ajk}
 + \sum_{ab}\nu^a_v \Phi_{1v}^{abb}
 +\sum_{ij} (\phi^{0i}_{v} - \phi^{0 i}_{v+\hat{e}_j} )e^{ji}_v
 \nonumber
\\
&& \left.
 + \sum_{ijk} \epsilon_{ijk} \phi^{0i}_{v} (\phi^{kj}_{n+1,v}
 + \phi^{0 k}_{v+\hat{e}_j}
 + \phi^{jk}_{v})
 -\sum_{ijk} \phi^{0 i}_{v+\hat{e}_j}
 \epsilon_{ijk} (\phi^{jk}_{v}
 + \phi^{jk}_{n+1,v})
 \right\} \label{lagexpandido}
\end{eqnarray}

Now that we have an explicit expression for the Lagrangian we can
proceed to identify the various terms. The theory has the
following Lagrange multipliers: $\phi^{0i}_{v}$ the ``vertical
component of the phase'' (which plays a role analogous to the time
component of the vector potential in Maxwell theory) and the
linearized lapse and shift. These quantities multiply times the
constraints of the linearized theory. Explicitly, the momentum and
Hamiltonian constraint read,
\begin{eqnarray}
C^a_v &=& \sum_b \Phi_{1v}^{abb} ,\label{mom}\\
C_v &=& \sum_{ijk} \epsilon_{ijk} \Phi_{1v}^{ijk}. \label{ham}
\end{eqnarray}

In order to get Gauss' law, we first take the variation of the
Lagrangian with respect to the Lagrange multiplier $\phi^{0i}_v$
to get,
\begin{equation}\label{Gauss1}
G^i_v \equiv \sum_a ( \,\xi^{ai}_{n+1,v} +
\xi^{\overline{a}i}_{n+1,v}\, ) = 0
\end{equation}
where
\begin{eqnarray}
\xi^{ai}_{n+1,v} &\equiv& +\delta^{ai} + e^{ai}_{n,v} + \sum_k
\epsilon_{aik}\, (-\phi^{ak}_{n,v} - \phi^{0 k}_{n,v+\hat{e}_a} -
\phi^{0k}_{n,v}+\phi^{ak}_{n+1,v})
\\
\xi^{\overline{a}i}_{n+1,v} &\equiv& -\delta^{ai} -
e^{ai}_{n,v-\hat{e}_a} + \sum_k \epsilon_{aik}\, (
\phi^{ak}_{n,v-\hat{e}_a} + \phi^{0 k}_{n,v} +
\phi^{0k}_{n,v-\hat{e}_a} + \phi^{ak}_{n+1,v-\hat{e}_a})
\end{eqnarray}

At the moment this does not appear to be a true constraint since
it involves variables at instant $n$ and at instant $n+1$. To
see that it actually is a constraint, we will call
$\xi^{ai}_{n+1,v}$ the component in the direction $\hat{e}_a$ of
a quantity that we will think of as an ``electric field'' (in the
sense that it is the quantity that satisfies the usual form of
Gauss law) and we will call 
$\xi^{\overline{a}i}_{n+1,v}$ the component in the direction
$-\hat{e}_a$, both at point $(n+1,v)$. To make
this more transparent, we need to see how they transform under
gauge transformations. To leading order the field  $e^a_{n,v}$ is
$e^a_{0,n,v} = \sum_i \delta^{ai}T^i$. We then define
\begin{equation}
\breve{e}^{\, a}_{n,v} = e^a_{n,v} +\frac{1}{4}[\, h^{0a}_{n,v}\,
e^a_{0,n,v}\, (h^{0a}_{n,v})^\dag - h^{a0}_{n,v}\, e^a_{0,n,v}\,
(h^{a0}_{n,v})^\dag \,] = e^a_{n,v} +\sum_{jk} \epsilon_{ajk}
\Phi_{1v}^{0ak} \, T^j\,,
\end{equation}
with the second equality valid up to second order. By inspection 
one sees that the field 
$\breve{e}^{\, a}_{n,v}$ is an element of the algebra that 
transforms like an electric field at  $(n,v)$
under gauge transformation. One can also show the following 
identities, valid to first order,
\begin{eqnarray}
\xi^{a}_{n+1,v} &\equiv& \sum_i \xi^{ai}_{n+1,v} \, T^i =
(h^0_{n,v})^\dag \, \breve{e}^{\, a}_{n,v} \, h^0_{n,v}
\\
\xi^{\overline{a}}_{n+1,v} &\equiv&
\sum_i\xi^{\overline{a}i}_{n+1,v} \, T^i = -
(h^a_{n+1,v-\hat{e}_a})^\dag \, \xi^{a}_{n+1,v-\hat{e}_a} \,
h^a_{n+1,v-\hat{e}_a}
\end{eqnarray}
from which one immediately sees that the quantities we identified
as components of the electric field have the appropriate
transformation properties under gauge transformations.

Therefore we can identify (\ref{Gauss1}) as the usual intuitive
expression of Gauss' law stating that field lines cannot emanate from
a point in vacuum.

We now turn our attention to the equations of motion. Given the
Lagrangian (\ref{lagexpandido}) we work out the equations of
motion from the canonical transformation. We start by computing
the canonical conjugate momentum to  $e$,
\begin{eqnarray}
P^{(e)ak}_{n+1,v}  &\equiv& \frac{\partial L(n,n+1)}{\partial
e^{ak}_{n+1,v}} = 0\,,
\\
P^{(e)ak}_{v}&\equiv& -\frac{\partial L(n,n+1)}{\partial
e^{ak}_{v}} = -\Phi^{a0k}_{1v} + 2 \sum_{ij} \epsilon^{kij}
\Phi^{aij}_{1v} =0.
\end{eqnarray}

Therefore the dynamics of $P^{(e)}$ is trivial. However, the last
equation can be viewed as an evolution equation for $\phi$ through
(\ref{FasePlaqueta}),
\begin{equation}\label{evol-phi}
\phi^{ak}_{n+1,v} = \phi^{ak}_{v} -\phi^{0k}_{v} +
\phi^{0k}_{v+\hat{e}_a}-2 \sum_{ij}\epsilon^{kij}
\Phi^{aij}_{1v}\,.
\end{equation}
Notice that by adding over indices $a$ belonging to a given plaquette
equation (\ref{evol-phi}), one effectively gets an evolution
equation for all the horizontal $\phi$'s in the plaquette. This is
due to the fact that the vertical contributions in
(\ref{FasePlaqueta}) will cancel out in pairs when adding through
the plaquette. Explicitly,
\begin{equation}\label{evol-Phi1}
\Phi^{abk}_{1,n+1,v} = \Phi^{abk}_{1v}
 -2 \sum_{ij}\epsilon^{kij} (\Phi^{aij}_{1v}
 + \Phi^{bij}_{1v+\hat{e}_a}
 - \Phi^{aij}_{1v+\hat{e}_b}
 - \Phi^{bij}_{1v})
\equiv \Phi^{abk}_{1v}
 -2 \sum_{ij}\epsilon^{kij} \sum_{d \in P_{ab}} \Phi^{dij}_{1v_d}
\end{equation}
where in the last term $v_d$ is the vertex in which the link $d$
originates and $P_{ab}$ is the plaquette spanned by $a$ and $b$.

We now consider the momentum canonically conjugate to $\phi$. We
start by computing the canonical conjugate momentum at instant
$n+1$
\begin{equation}\label{evol-pi-phi-(n+1)}
P^{(\phi)ak}_{n+1,v} \equiv \frac{\partial L(n,n+1)}{\partial
\phi^{ak}_{n+1,v}} = -e^{ak}_v + \sum_{i}\epsilon^{aki}(
\phi^{ai}_v + \phi^{0i}_v + \phi^{0i}_{v+\hat{e}_a})\,.
\end{equation}
The momentum can be written in terms of the electric field in 
an expression that sees parallels  the usual relation between
the electric field and the canonical momentum in the lattice
$\xi^{ak}$ as,
\begin{equation}
P^{(\phi)ak}_{n+1,v} = \delta_{ak}- \xi^{ak}_{n+1,v} +
\sum_i\epsilon_{aki}\, \phi^{ai}_{n+1,v}
\end{equation}
and in terms of it, Gauss' law (\ref{Gauss1}) can be written as,
\begin{equation}\label{Gauss2}
G^k_{n+1,v} =
 - \sum_a \left(P^{(\phi)ak}_{n+1,v}
 - P^{(\phi)ak}_{n+1,v-\hat{e}_a}
 - \sum_i \epsilon^{aki}(\phi^{ai}_{n+1,v}+\phi^{ai}_{n+1,v-\hat{e}_a})
 \right)\,.
\end{equation}
This final expression for Gauss' law is a genuine constraint, in the
sense that all variables are expressed at the same instant of time.

We now compute the 
momentum conjugate to  $\phi$ at instant $n$,
\begin{eqnarray}\label{evol-pi-phi-n}
  P^{(\phi)ab}_v &=&
  - \frac{\partial L(n,n+1)}
  {\partial \phi^{ab}_v} =
  - e^{ab}_v
  + \nu^a_v
  - \nu^a_{v-\hat{e}_b}
  + \delta_{ab}
  \sum_i\left(- \nu^i_v + \nu^i_{v-\hat{e}_i}\right)
\\
  &&+ 2\left(- \phi^{aa}_{v-\hat{e}_b}
  + \phi^{aa}_{v+\hat{e}_b}
  + \phi^{ba}_v
  - \phi^{ba}_{v+\hat{e}_a}
  - \phi^{ba}_{v-\hat{e}_b}
  - \phi^{ba}_{v+\hat{e}_a-\hat{e}_b}\right)\nonumber
\\
  &&+ 2\delta_{ab}\sum_i
  \left(\phi^{a i}_{v-\hat{e}_i}
  - \phi^{a i}_{v+\hat{e}_i}
  - \phi^{ii}_v
  + \phi^{ii}_{v+\hat{e}_a}
  + \phi^{ii}_{v-\hat{e}_i}
  + \phi^{ii}_{v+\hat{e}_a-\hat{e}_i}\right)\nonumber
\\
  &&+ \sum_i\epsilon_{abi}
  \left( -2\nu_v
  + 2\nu_{v-\hat{e}_i}
  + \phi^{0i}_v
  - \phi^{0i}_{v+\hat{e}_a}
  - 2e^{aa}_v
  + 2e^{aa}_{v-\hat{e}_i}
  - 2e^{b i}_v
  + 2e^{b i}_{v-\hat{e}_b}\right.\nonumber
\\
  &&\left.
  - 2e^{ii}_v
  + 2e^{ii}_{v-\hat{e}_i}
  + \phi^{a i}_{n+1,v} \right)
  + 2\delta_{ab}\sum_{i,j}
  \epsilon_{aij}
  \left(e^{a i}_v
  - e^{a i}_{v-\hat{e}_j}\right).\nonumber
\end{eqnarray}

One still needs to replace the expressions for the $e$'s and for the
$\phi$'s evaluated at instant $n+1$.  The resulting substitutions lead
to lengthy expressions that are not particularly illuminating, and will
not be needed in what follows, so we will not display them here.  We
point out however, that the resulting scheme is not an explicit one for
the $P^{(\phi)}$'s.  Since these variables do not arise in the
constraints, we do not need this evolution equation to show mimetism.

One now needs to show that the evolution is mimetic, that is, it
preserves the discrete constraints (\ref{mom}, \ref{ham},
\ref{Gauss2}). Using the evolution equation (\ref{evol-Phi1})
one gets that,
\begin{eqnarray}
C^a_{n+1,v} &=& C^a_v + C_v - C_{v+\hat{e}_a} ,\\
C_{n+1,v} &=& C_v + 4 \sum_a ( C^a_{v+\hat{e}_a} - C^a_{v}) .
\end{eqnarray}

To study the time evolution of Gauss' law one needs equations
(\ref{evol-phi},\ref{evol-pi-phi-(n+1)},\ref{evol-pi-phi-n}) and
one gets that 
\begin{equation}
G^k_{n+1,v} = G^k_{n,v}\,.
\end{equation}

As in the previous section, we have checked these identities using
computer algebra.

\section{Discussion and conclusions}

The consistent discretization scheme is such that it yields
a set of discrete equations for the evolution equations and
the constraints of general relativity that is compatible, that
is, all can be solved simultaneously. It does so at the price
of determining the Lagrange multipliers (the lapse and the
shift). In the linearized case we have shown that one can
discretize the theory in such a way that the Lagrange multipliers
are not determined and nevertheless the theory is consistent.

When one discretizes a theory there is always an ambiguity in how to
proceed. Among the ambiguities we have the dependence on how one
chooses to represent the derivatives. What we have found is that in
the linearized case one can choose certain derivative operators for
which the Lagrange multipliers are not determined. It should be noted
that the consistent discretization scheme would work even if one did
not choose the derivatives this way, but the Lagrange multipliers will
be determined in order to have a consistent set of equations. This is
true both in the case of the traditional variables and also in the
Ashtekar variables. In the latter case there is an additional element
that is the presence of an extra constraint: the Gauss law. We have
also chosen a specific way of discretizing the theory in such a way
that Gauss' law is implemented exactly in the discrete theory (this is
standard in Yang--Mills theory on the lattice, and implies that the
discrete formulation is gauge invariant, and also that these
discretizations are mimetic, though this is rarely emphasized in the
Yang--Mills literature). In the case of the traditional variables, one
can also associate mimetism with gauge invariance. The action we chose
to work with is invariant under linearized coordinate transformations
of the form $h'_{\mu\nu} = h_{\mu\nu} +\xi_{(\mu,\nu)}$. The discrete
action, if one chooses a derivative operator such that the second
derivatives coincide with the derivative of a first derivative and
satisfies summation by parts, is invariant under a discrete version of
the above symmetry.  This symmetry is generated canonically by the
discrete constraints. This explains in a geometrically nice way why
mimetism was possible in the linearized case.

In the case of Lorentzian general relativity written in terms of
Ashtekar's variables, the presence of the terms $(\Gamma_a^i-\sigma
\beta A_a^i)$ in the Hamiltonian make it more difficult to discretize
the action in such a way that the Gauss law is preserved exactly. This
is because the $\Gamma_a^i$'s have to be written in terms of the
triads and the resulting expressions are not easy to discretize on the
lattice preserving the internal symmetry (unlike the $A_a^i$'s which
are readily discretized by considering a parallel transport operator
along the elementary links). It may be possible using dual lattices to
discretize such terms in an invariant way, but this will require
further study. There is no problem applying our discretization
technique to this case directly, but what will happen is that the 
internal symmetry will be broken, the Lagrange multipliers associated
with the Gauss law will be determined and the resulting discretization
will not be mimetic in the traditional sense of the word. It is clear
that further work is needed before this kind of discretizations
will be useful numerically.

In the full nonlinear case either with the traditional or the new
variables, the constraints involve both coordinates and momenta and
therefore the application of our technique will determine the Lagrange
multipliers and will therefore not furnish a mimetic discretization in
the traditional sense.  The resulting discrete theory is consistent,
but it does so at the expense of determining the Lagrange
multipliers. Based on what we learned from the linearized case, we can
conclude that the only remaining possibility for a formulation that is
mimetic in the traditional sense would be to implement the symmetries
implied by the constraints exactly in the lattice. Since the
symmetries implied by the diffeomorphism constraint are broken by the
introduction of the lattice it appears unlikely that such a
formulation would ever be found.

The conclusion we can draw from this is that for the case of full
nonlinear general relativity, the closest one can come to a
formulation that preserves the constraints under evolution is the
proposal of consistent discretizations we have introduced. Such
proposal is not mimetic in the traditional sense in that it imposes
the constraints by determining the Lagrange multipliers. This proposal
has many new aspects that are currently in investigation. It has been
successfully applied in cosmological examples and is now being studied
in detail for the Gowdy space-times. If it works for this example, it
is likely that it could be applied successfully in general, but this
obviously requires further study.

Something to be noticed is that it is not clear that the formulations
we presented are going to be useful numerically. In particular,
the fact that they are not based on manifestly hyperbolic equations.
We have presented a first step towards showing stability of the
scheme in a particular situation, but a fuller analysis should be
carried out to determine if the scheme is stable in general.
Numerical relativity codes also use more sophisticated time stepping
techniques than the one we use. It is clear, however, that our 
method can accommodate more elaborate discretizations of the time
derivatives and the calculations in this paper could be repeated
in that case. Another interesting point would be to attempt to 
apply the techniques in this paper to the several manifestly 
hyperbolic formulations of the Einstein equations that have 
been proposed in the last few years. Unfortunately, few of them
have been worked out in the context of an action principle, but
this difficulty could presumably be remedied. This would also 
allow to study within our framework manifolds with boundaries.

Another element of interest is the impact of the choice of the
derivative operators on the construction of consistent discrete
theories. The consistent technique will work no matter what 
derivative operators are chosen. But here we have learnt that
one can choose them in such a way that the linearized theory
is automatically mimetic. We would like to argue that the
level of accuracy with which the consistent discretization
enforces the constraints is improved when one chooses a 
formulation that is mimetic at the linear level, at least for
weak fields. The argument is simple. In the consistent discretization
scheme the constraints that are enforced exactly have the
form $\phi(q_n,p_{n+1})=0$. The constraints one would like
to see enforced are of the form $\phi(q_n,p_n)=0$. Starting
from the former, and using the equations of motion one has
that $\phi(q_n,p_n+O(p^2))=0$ where the terms that correct
$p_n$ are of order $p^2$ (or $q^2$ or mixed but quadratic). This
is true if the theory is mimetic in the linearized level.
Otherwise one would have $\phi(q_n,p_n+O(p))=0$. Therefore
choosing a discretization that is mimetic at the linearized level,
at least for weak fields, implies that the constraints are
tracked more accurately in the full non-linear theory when
one discretizes consistently.

Summarizing, we have shown that the consistent discretization scheme
we have introduced recently, when applied to 
general relativity  discretized around Minkowski spacetime  yields a
formulation that is mimetic. That is, a formulation in which the
discrete constraints are exactly preserved upon discrete evolution. 
We have also argued that for the full nonlinear case, the use
of consistent discretizations appears as a possibility to yield
a formulation that is close to the intention of mimetic formulations,
although only approximately.

\section{Acknowledgments}
We wish to thank Manuel Tiglio for discussions and Luis Lehner
and Olivier Sarbach for comments on the manuscript.
This work was supported by grant NSF-PHY0244335, NASA-NAG5-13430,
DID-USB grant GID-30, Fonacit grant G-2001000712
and
funds from the Horace Hearne Jr. Laboratory for Theoretical Physics
and the Abdus Salam International Center for Theoretical Physics.

\end{document}